\documentclass[fleqn,aps,prl,nofootinbib,twocolumn,letterpaper,superscriptaddress]{revtex4}
\usepackage{dcolumn}
\usepackage{color}
\usepackage{graphicx}
\usepackage[colorlinks=true]{hyperref}
\usepackage{amsmath}
\usepackage{multirow}
\usepackage{bm}
\usepackage{float}
\usepackage{lineno}
\usepackage{caption}
\begin{document}

\title{A search for solar axions and anomalous neutrino magnetic moment with the complete PandaX-II data}
\date{November 20, 2020}
\begin{abstract}
We report a search for new physics signals using the low energy electron recoil events in the complete data set from PandaX-II, in light of the recent event excess reported by XENON1T. The data 
correspond to a total exposure of 100.7 ton-day with liquid xenon. 
  With robust estimates of the dominant background spectra, we perform sensitive searches on solar axions and neutrinos with enhanced magnetic moment. We find that the axion-electron coupling $g_{Ae}<4.6\times 10^{-12}$ for an axion mass less than $\rm 0.1~keV/c^2$
  and the neutrino magnetic moment $\mu_{\nu}<4.9\times 10^{-11}\mu_{B}$ at 90\% confidence level. The observed excess from XENON1T is within our experimental constraints. 
\end{abstract}

\def\shKeyLab{INPAC, School of Physics and Astronomy, Shanghai Jiao Tong University, Shanghai Key Laboratory for Particle Physics and Cosmology, Shanghai 200240, China}
\def\BUAA{School of Physics, Beihang University, Beijing 100191, China}
\def\BUAALab{International Research Center for Nuclei and Particles in the Cosmos \& Beijing Key Laboratory of Advanced Nuclear Materials and Physics, Beihang University, Beijing 100191, China}
\def\pku{School of Physics, Peking University, Beijing 100871, China}
\def\YaLongSD{Yalong River Hydropower Development Company, Ltd., 288 Shuanglin Road, Chengdu 610051, China}
\def\IAP{Shanghai Institute of Applied Physics, Chinese Academy of Sciences, 201800 Shanghai, China}
\def\CHEPpku{Center for High Energy Physics, Peking University, Beijing 100871, China}
\def\SDU{School of Physics and Key Laboratory of Particle Physics and Particle Irradiation (MOE), Shandong University, Jinan 250100, China}
\def\UMD{Department of Physics, University of Maryland, College Park, Maryland 20742, USA}
\def\TDLee{Tsung-Dao Lee institute, Shanghai Jiao Tong University, Shanghai, 200240, China}
\def\MESJTU{School of Mechanical Engineering, Shanghai Jiao Tong University, Shanghai 200240, China}
\def\SYU{School of Physics, Sun Yat-Sen University, Guangzhou 510275, China}
\def\NKU{School of Physics, Nankai University, Tianjin 300071, China}
\def\FDU{Key Laboratory of Nuclear Physics and Ion-beam Application (MOE), Institute of Modern Physics, Fudan University, Shanghai 200433, China}
\def\USST{School of Medical Instrument and Food Engineering, University of Shanghai for Science and Technology, Shanghai 200093, China}
\def\SJTUSC{Shanghai Jiao Tong University Sichuan Research Institute, Chengdu 610213, China}
\def\Princeton{Physics Department, Princeton University, Princeton, NJ 08544, USA}
\def\MIT{Department of Physics, Massachusetts Institute of Technology, Cambridge, MA 02139, USA}
\def\SARI{Shanghai Advanced Research Institute, Chinese Academy of Sciences, Shanghai 201210, China}

\affiliation{\shKeyLab}
\author{Xiaopeng Zhou}\affiliation{\BUAA}
\author{Xinning Zeng}   
\author{Xuyang Ning}
\author{Abdusalam Abdukerim}
\author{Wei Chen}\affiliation{\shKeyLab}
\author{Xun Chen}\affiliation{\shKeyLab}\affiliation{\SJTUSC}
\author{Yunhua Chen}\affiliation{\YaLongSD}
\author{Chen Cheng}\affiliation{\SYU}
\author{Xiangyi Cui}\affiliation{\TDLee}
\author{Yingjie Fan}\affiliation{\NKU}
\author{Deqing Fang}
\author{Changbo Fu}\affiliation{\FDU}
\author{Mengting Fu}\affiliation{\pku}
\author{Lisheng Geng}\affiliation{\BUAA}\affiliation{\BUAALab}
\author{Karl Giboni}
\author{Linhui Gu}\affiliation{\shKeyLab}
\author{Xuyuan Guo}\affiliation{\YaLongSD}
\author{Ke Han}\email[Corresponding author: ]{ke.han@sjtu.edu.cn}
\author{Changda He}\affiliation{\shKeyLab}
\author{Shengming He}\affiliation{\YaLongSD}
\author{Di Huang}\affiliation{\shKeyLab}
\author{Yan Huang}\affiliation{\YaLongSD}
\author{Yanlin Huang}\affiliation{\USST}
\author{Zhou Huang}\affiliation{\shKeyLab}
\author{Xiangdong Ji}\affiliation{\UMD}\affiliation{\TDLee}
\author{Yonglin Ju}\affiliation{\MESJTU}
\author{Shuaijie Li}\affiliation{\TDLee}
\author{Huaxuan Liu}\affiliation{\MESJTU}
\author{Jianglai Liu}\email[Spokesperson: ]{jianglai.liu@sjtu.edu.cn}\affiliation{\shKeyLab}\affiliation{\TDLee}\affiliation{\SJTUSC}
\author{Xiaoying Lu}\affiliation{\SDU}
\author{Wenbo Ma}\affiliation{\shKeyLab}
\author{Yugang Ma}\affiliation{\FDU}
\author{Yajun Mao}\affiliation{\pku}
\author{Yue Meng}\affiliation{\shKeyLab}\affiliation{\SJTUSC}
\author{Kaixiang Ni}\affiliation{\shKeyLab}
\author{Jinhua Ning}\affiliation{\YaLongSD}
\author{Xiangxiang Ren}\affiliation{\SDU}
\author{Changsong Shang}\affiliation{\YaLongSD}
\author{Guofang Shen}\affiliation{\BUAA}
\author{Lin Si}\affiliation{\shKeyLab}
\author{Andi Tan}
\affiliation{\UMD}
\author{Anqing Wang}\affiliation{\SDU}
\author{Hongwei Wang}\affiliation{\IAP}\affiliation{\SARI}
\author{Meng Wang}\affiliation{\SDU}
\author{Qiuhong Wang}\affiliation{\IAP}
\author{Siguang Wang}\affiliation{\pku}
\author{Wei Wang}\affiliation{\SYU}
\author{Xiuli Wang}\affiliation{\MESJTU}
\author{Zhou Wang}\affiliation{\shKeyLab}\affiliation{\SJTUSC}
\author{Mengmeng Wu}\affiliation{\SYU}
\author{Shiyong Wu}\affiliation{\YaLongSD}
\author{Weihao Wu}
\author{Jingkai Xia}\affiliation{\shKeyLab}
\author{Mengjiao Xiao}
\affiliation{\UMD}
\author{Pengwei Xie}\affiliation{\TDLee}
\author{Binbin Yan}\affiliation{\shKeyLab}
\author{Jijun Yang}
\author{Yong Yang}\affiliation{\shKeyLab}
\author{Chunxu Yu}\affiliation{\NKU}
\author{Jumin Yuan}\affiliation{\SDU}
\author{Ying Yuan}\affiliation{\shKeyLab}
\author{Dan Zhang}\affiliation{\UMD}
\author{Tao Zhang}
\author{Li Zhao}\affiliation{\shKeyLab}
\author{Qibin Zheng}\affiliation{\USST}
\author{Jifang Zhou}\affiliation{\YaLongSD}
\author{Ning Zhou}
\email[Corresponding author: ]{nzhou@sjtu.edu.cn}\affiliation{\shKeyLab}

\collaboration{PandaX-II Collaboration}
\noaffiliation
\maketitle

Dark matter particles have been searched extensively in underground low background experiments in the past a few decades~\cite{Gaitskell:2004gd, Undagoitia:2015gya, NaturePhysics}. Nuclear recoil (NR) signals from dark matter-nucleus collisions, with a typical energy
scale of 10 keV, are mixed primarily with electron recoil (ER) background due to gamma or beta radioactivity. Many experimental
techniques have been developed to suppress the ER background,
and to enhance the capability to identify NR signals. At the same time, utilizing the ER events to search for new physics signals such as
dark matter-electron scattering~\cite{PhysRevLett.109.021301,PhysRevD.96.043017,PhysRevLett.123.251801,PhysRevLett.121.111303darkside,PhysRevLett.123.181802,Emken_2019,Abramoff:2019dfb}, axion-electron interactions~\cite{Redondo:2013jcap,Axioelectric-PRD1987,Fe57-Axion-PRL1995,Primkaoff},
and neutrino-electron scattering via neutrino magnetic
moment~\cite{fujikawa,magmu, Billard:2014prd, Huang:2019jcap} also becomes increasingly appealing.

Recently, XENON1T released the energy spectrum of ER
events in their dark matter search data~\cite{Xenon1Texcess}. With an exposure of 0.65 ton-year with liquid xenon target, an excess of about $53\pm 15$ events is identified between 1 to 7 keV above expected background. They report that such an excess is consistent with a non-zero solar axion-electron coupling constant $g_{Ae}$ at $3.5~\sigma$, or solar neutrino-electron scattering with anomalously large magnetic moment at $3.2~\sigma$. However, the excess could also be explained by a trace amount of unexpected tritium in the detector. In this paper, we report an independent investigation of the low energy ER data from the full exposure of PandaX-II, where the spectra of dominant background components are well controlled by direct measurement or calibration.

PandaX-II is a 580-kg dual-phase liquid xenon experiment~\cite{PandaX-PRL-2016}, operating in the China Jinping Underground Laboratory~\cite{Kang_2010}. The setup of the
experiment has been reported previously~\cite{Tan:2016diz,Cui:2017nnn,PandaX-PRL-2016}, so only essential information
relevant to this analysis is provided here. The PandaX-II target is a
cylindrical-shaped time projection chamber. Liquid xenon is continuously purified through two circulation loops with a total mass flow rate of about 560 kg/day through hot getters~\cite{SAES}.
The photons from the prompt scintillation ($S1$)
and delayed electroluminescence photons from ionized electrons ($S2$)
are detected by a top and bottom array of 3-inch photomultiplier tubes (PMTs). Single-scatter
events are selected with a requirement of one $S1$ and one $S2$ signal based on which the three-dimensional position of the interaction can be reconstructed.
The energy of the event is reconstructed by a linear combination of $S1$ and $S2$ as, $E\rm_{rec}$ =
13.7 eV ($S1$/PDE + $S2$/EEE/SEG), in which the PDE, EEE, and SEG refer to the
average photon detection efficiency, electron extraction efficiency,
and single electron gain, respectively, and can be calibrated using
full absorption ER peaks.

The complete data sets, including Run 9, Run 10, and Run 11, are used~\cite{Wang:2020cpc}. In between Run 9 and Run 10 in summer 2016, an ER calibration campaign was carried out with tritiated methane injection (named T1), with a peak tritium rate of 1~Bq, corresponding to a $2.1\times10^{-19}$~mol/mol of tritium concentration in the target. After the calibration, attempts were made to remove the methane through the hot getter, an effective approach demonstrated by the LUX collaboration~\cite{Akerib:2015wdi}. However, after the
initial decrease, the tritium rate plateaued in the detector to about 1 mBq.
In order to reduce this background, xenon from the detector was recuperated and distilled in spring 2017, during which the detector was completely warmed up and flushed with warm xenon through the hot getters. Run 10 was resumed after the distillation campaign, and a significant decrease of tritium level was observed~\cite{Cui:2017nnn}. Afterwards, an extended data taking (Run 11) was carried out, but divided into two spans (11-1 and 11-2) due to an unexpected air leakage in between leading to different background levels~\cite{Wang:2020cpc}. Instead of tritium, we injected short-lived $^{220}$Rn into the detector for ER calibration. After Run 11, another tritium calibration was performed (T2). The two tritium
injections, separated by three years, allow us to make an accurate measurement of its spectrum, which eliminates potential systematic uncertainty in spectral modeling in this analysis.

The low-level analysis in this work follows the same procedure as that in Ref.~\cite{Cui:2017nnn}. The data quality cuts (determined blindly) in the main dark matter analysis are adopted, although the range of energy is widened (see later). Detector parameters PDE, EEE, and SEG are taken from Ref.~\cite{Wang:2020cpc}, which are consistent among the three data sets within $10\%$ (fractional).

\begin{figure}[bt]
  \centering
  \includegraphics[width=\columnwidth]{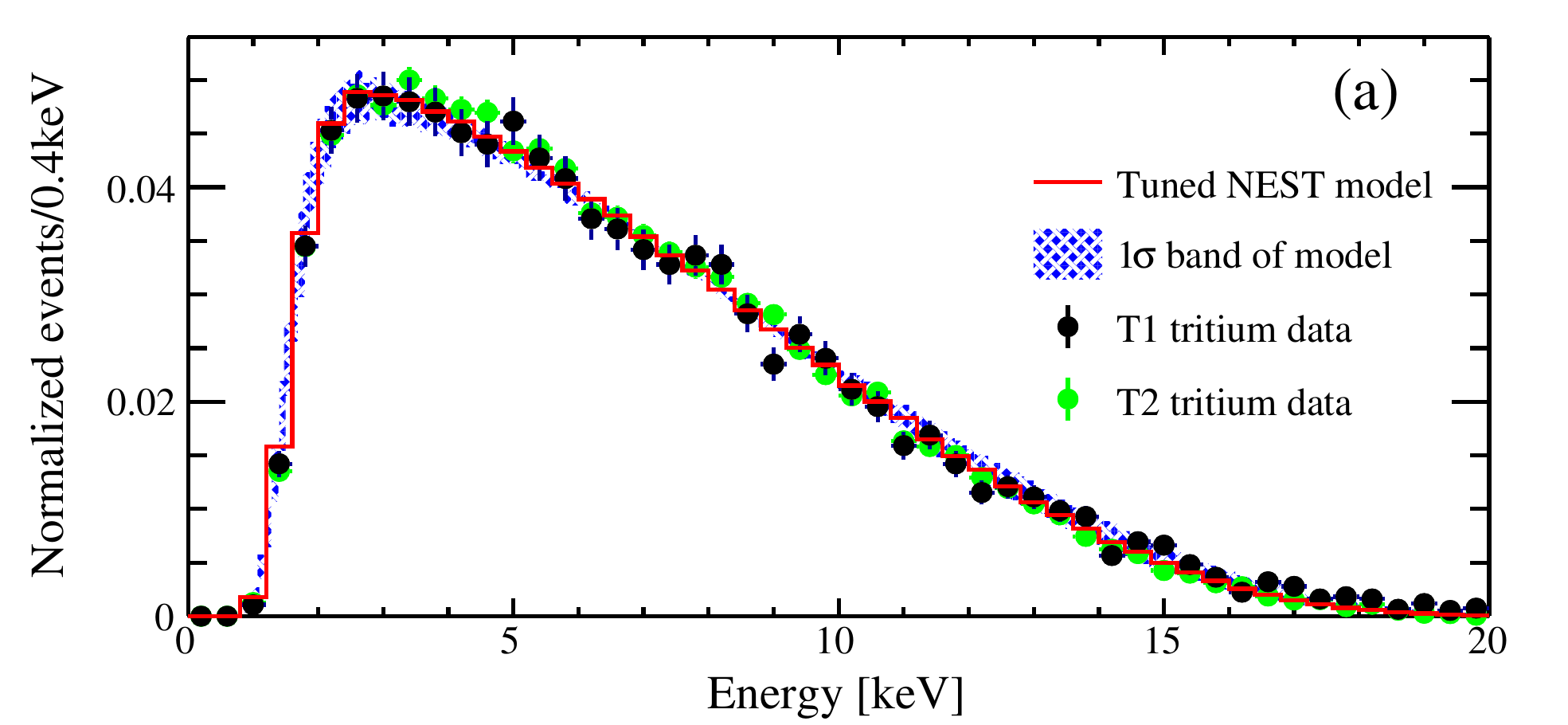}
  \includegraphics[width=\columnwidth]{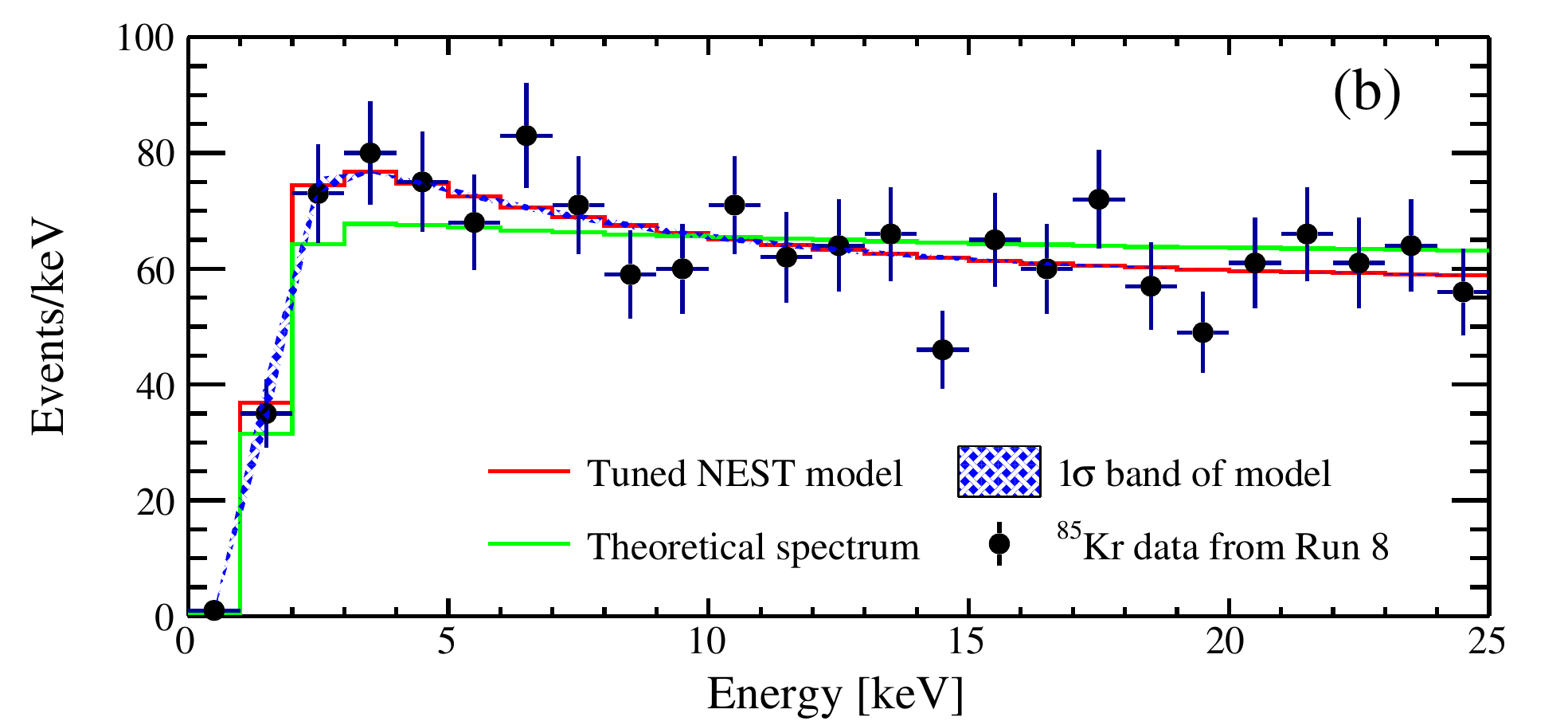}
  \includegraphics[width=\columnwidth]{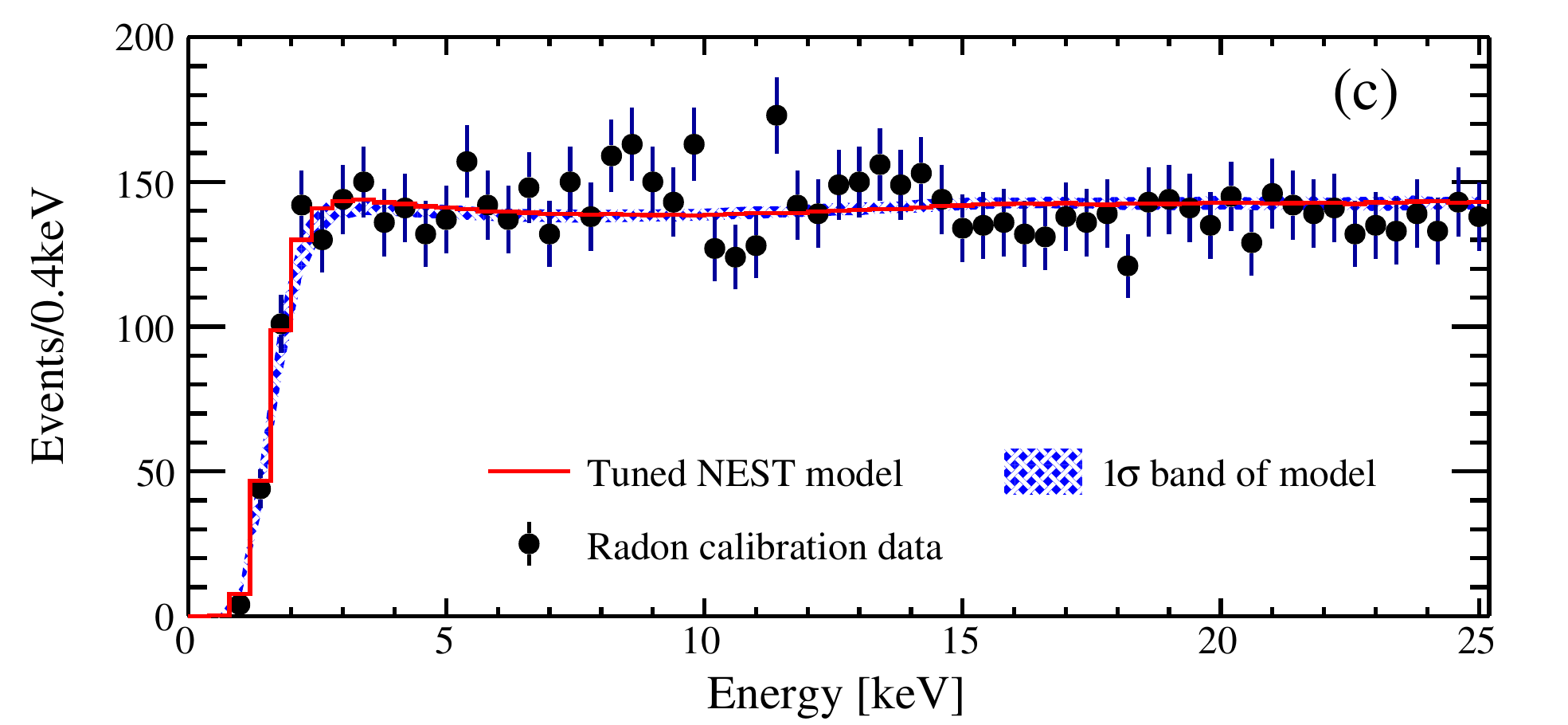}
  \caption{(a) BLS-corrected tritium energy spectra from T1 and T2, and the NEST model; (b)
  measured spectrum of $^{85}$Kr, our best fit, and a recent theoretical evaluation; (c) measured $^{220}$Rn calibration data in comparison with the NEST model. The shaded area represents systematic uncertainty in spectrum shape due to PDE ($4.9\%$ rel. uncert.), EEE$\times$SEG($4.8\%$ rel. uncert.), and the BLS corrections. }
  \label{fig:bkgdspectrum}
\end{figure}

To select the final set of ER events, the radius cut is modified from $r^2<720~\rm cm^2$ in the dark matter search~\cite{Wang:2020cpc} to $550~\rm cm^2$ 
to suppress the so-called surface background to less than one event.
The corresponding fiducial mass is 251 kg. Also different from the main dark matter analysis, the energy region is extended to 25 keV in order to have better control of the shape of the beta-emitting background in the detector. 
Within this range, the tritium, $^{85}$Kr, material radioactivity and $^{222}$Rn are the four major background components. The expected energy spectra from the material, $^{222}$Rn, and solar neutrinos are mostly flat below 25 keV. For simplicity and to avoid degeneracy in the spectral fit, they are combined into a single component, named ``flat ER'' background in short. The residual background including
$\rm ^{136}Xe$, accidental coincidence of isolated $S1$s and $S2$s, and neutron events are all taken from the dark matter analysis but with updated fiducial volume cut.
Contributions of $^{39}$Ar and $^{37}$Ar are estimated to be negligible in these data sets.

The critical ingredient of this analysis is to have a robust estimate of the background spectra. At low energy, a subtle instrumental non-linearity arises from the baseline suppression (BLS) threshold which introduces channel-wise signal inefficiency, particularly for PMTs operated under low gains. As a result, both $S1$ and $S2$ are subject to suppression factors~\cite{Wang:2020cpc}, leading to a nonlinear compression of the spectrum and apparent excess
of events towards the low end. A special calibration was carried out to measure the two suppression factors directly at different PMT gain settings~\cite{Wang:2020cpc}, so the BLS effects can be properly corrected for the entire data set. This is particularly important in our understanding of the tritium spectrum, as its shape could be distorted more acutely. 
The validity of the BLS correction is demonstrated in Fig.~\ref{fig:bkgdspectrum}a), where a comparison is made on tritium energy spectra in T1 and T2, corrected for their corresponding BLS effects. The two spectra agree with each other with $\rm \chi^2/NDF= 69.4/50$. The measured spectra are also in good agreement with the tuned NEST2.0 model~\cite{szydagis_m_2018_1314669}, with parameters identical to that used in the dark matter analysis~\cite{Wang:2020cpc}.

The spectrum of $\rm ^{85}Kr$ background is measured directly using our commissioning data sets (Run 8), where a high $\rm ^{85}Kr$ concentration is identified and contributes to more than $98\%$ of the low energy ER events~\cite{Tan:2016diz}. The shape of $^{85}$Kr is extracted by fitting the data with an exponential function, as shown in Fig.~\ref{fig:bkgdspectrum}b). A recent theoretical calculation~\cite{theorybetadecay} is compared with the data, where a sizable difference is observed, indicating potential systematics from both ends. In this analysis, the difference is conservatively taken as the shape uncertainty of $\rm ^{85}Kr$. 

The shape of the ``flat ER'' background is studied with the $^{220}$Rn
injection data~\cite{Ma:2020kll}. For comparison, using the ER model in Ref.~\cite{Wang:2020cpc} with a flat input energy spectrum, the resulting $E\rm_{rec}$ is in good agreement with the data ($\chi^2$/NDF = 48.7/63), as shown in Fig.~\ref{fig:bkgdspectrum}c). Theoretical shape uncertainties of the ``flat ER'' components including $^{214}$Pb ~\cite{Xenon1Texcess,theorybetadecay} are taken into account, which are of up to a few percent level, and have less than 1\% impact on the final spectrum fit.

In total 2111 events survive after all cuts, with 646, 249, 382, and 834 events in 
Run 9 (20.0 ton-day), Run 10 (19.4 ton-day), Run 11-1 (24.2 ton-day), and Run 11-2 (37.1 ton-day). 
With tightened fiducial volume cut, we omit the position dependence in this analysis and generate background and signal probability density functions in
two-dimensional space of $S1$ and $S2$. 

\begin{figure}[tb]
\centering
\includegraphics[width=\columnwidth]{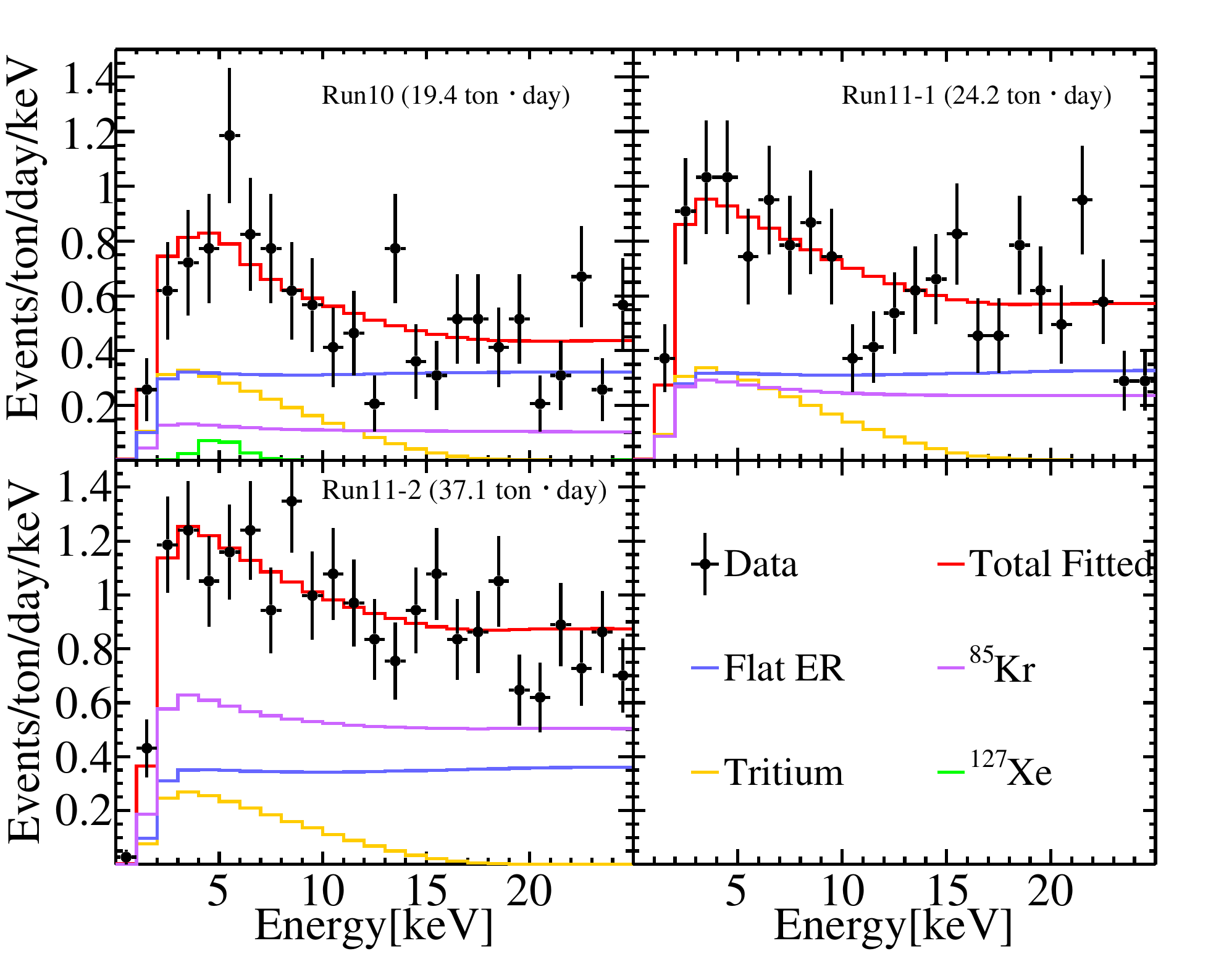}
\caption{Electron recoil energy distributions for Runs 10, 11-1, and 11-2 with background-only pre-fits. Likelihood fits are performed in two-dimensional space. The background due to $^{136}$Xe, neutron, and accidentals is not drawn in the figure.
}
\label{fig:fit}
\end{figure}

An unbinned likelihood fit is performed to test the background and signal hypotheses, where the construction of likelihood function is identical to that in Ref.~\cite{Cui:2017nnn}. 
In Run 10 and Run 11, to estimate the tritium contribution, a background-only pre-fit is performed independently for each run (span). The fit results are shown in Fig.~\ref{fig:fit}. 
The resulting tritium rates are $\rm 0.041 \pm 0.013$, $\rm 0.043\pm 0.014$, and
$\rm 0.035\pm 0.019~\mu Bq/kg$ for Runs 10, 11-1, and 11-2, consistent with a constant tritium decay rate
where the statistical uncertainty is dominant. 
Another fit is performed with a common tritium normalization in the runs. 
The best fit tritium rate is $\rm 0.040\pm 0.010~(stat.+sys.)~\mu Bq/kg$, translating into a concentration of $(4.9\pm1.2) \times 10^{-24}$~mol/mol in xenon. 
Similar fitting test is performed with Run 9 data, and the result is consistent with the tritium-free scenario as expected.
Therefore, in the signal hypothesis test discussed below, tritium background is not considered in Run 9 and the overall tritium normalization in Run 10 and Run 11 is floating in the fit. 

\begin{figure}[hbtp]
  \centering
  \includegraphics[width=0.44\textwidth]{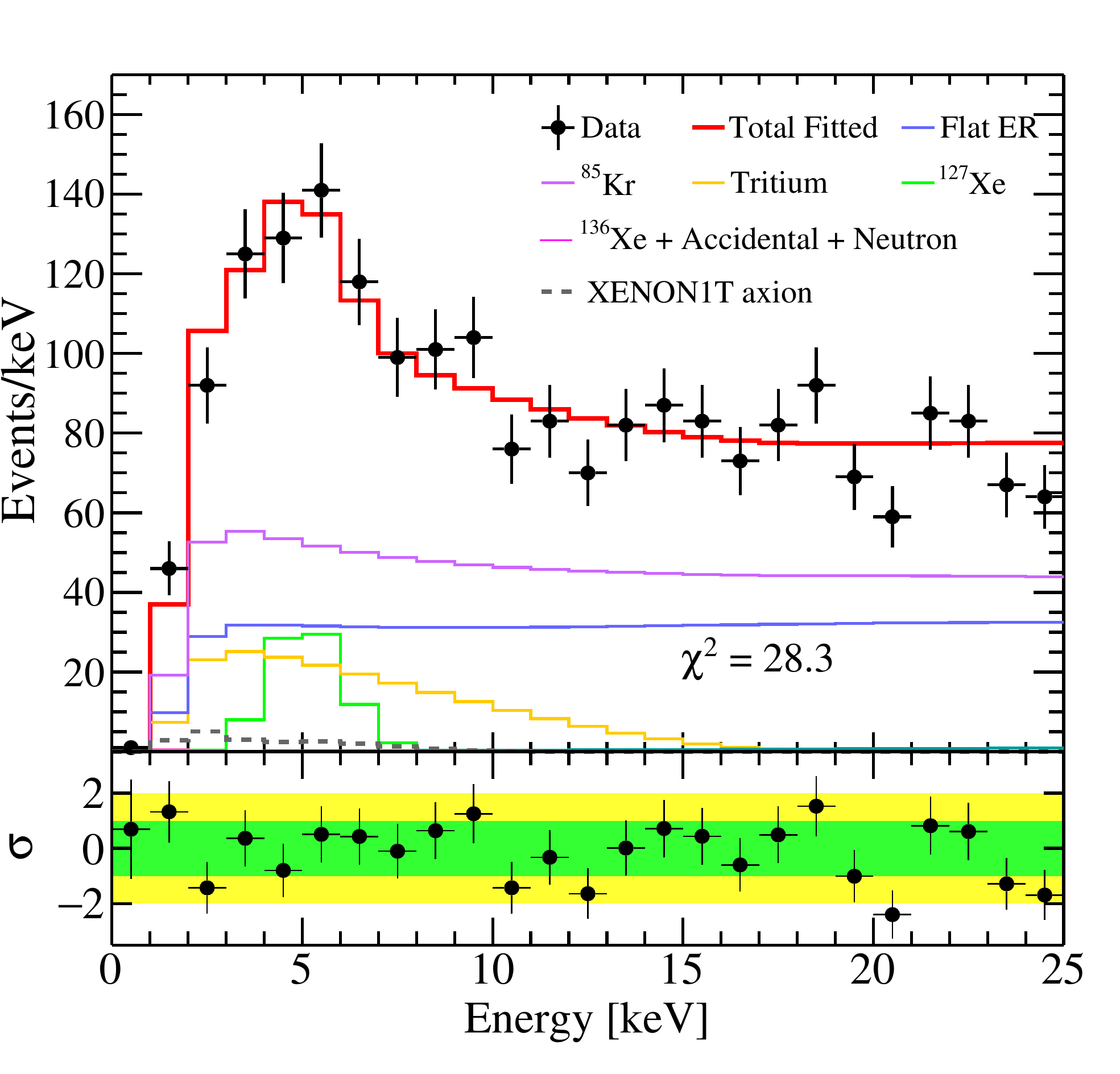}
  \caption{Low energy spectrum of electron recoil events for the total 100.7 ton-day data. Simultaneous best-fit background contributions are overlaid, where tritium background rate is treated as the same in Run 10 and Run 11. The expected axion signal with XENON1T best-fit signal strength is shown as a dashed grey line.}
  \label{fig:fitall}
\end{figure}

Table~\ref{tab:f} summarizes the background composition from the background-only fit. The summed energy spectrum from all runs is shown in Fig.~\ref{fig:fitall}, with best-fit background contributions superposed. 
The data are consistent within $1\sigma$ fluctuation of the background-only hypothesis. 

\begin{table}[hbtp]
  \centering
  \begin{tabular}{ccccc}\\\hline
  Events  & Run 9 & Run 10 & Run 11-1 & Run 11-2\\ \hline 
  $^{127}$Xe & 77.3 & 3.5 & 0.0 & 0.0 \\ \hline
  tritium & 0.0 & 49.6 & 60.1 & 92.2 \\\hline
  $^{85}$Kr & 418.2 & 51.1 & 146.0 & 479.7 \\\hline
  flat ER & 148.1 & 143.6 & 176.1 & 270.1 \\\hline
  accidental & 1.5 & 0.8 & 0.8 & 1.2 \\\hline
  neutron & 0.6 & 0.4 & 0.5 & 0.8 \\\hline
  $^{136}$Xe & 2.3 & 2.2 & 2.7 & 4.1 \\ \hline
  Total & $648.1 \pm 35.3$  & $251.2 \pm 22.1$  & $386.1 \pm 32.5$   & $848.1 \pm 52.7$ \\ \hline\hline
  Data & 646 & 249 & 382 & 834 \\\hline
  \end{tabular}
  \caption{Summary of the best fit background values and data from the background-only likelihood fit.}
  \label{tab:f}
\end{table}

Based on the above, we perform tests on the axion and neutrino magnetic momentum hypotheses with our data. For the axion hypothesis, we consider the Atomic recombination and de-excitation, Bremsstrahlung and Compton (ABC) solar axion model~\cite{Redondo:2013jcap}. 
The best fit axion signal yields 15.8 events with statistical-plus-systematic uncertainty band $[0,84.8]$.
Assuming XENON1T best fit signal strength~\cite{Wang:2020cpc} ($g_{Ae}=3.15\times10^{-12}$ for axion mass smaller than 
$0.1~{\rm keV}/c^2$), the expected number of signals would be 20.4 events in PandaX-II. Therefore, our data is compatible with XENON1T excess within 1$\sigma$ in number of events, but is also consistent with background fluctuations. 

\begin{figure}[H]
  \centering
  \includegraphics[width=0.85\columnwidth]{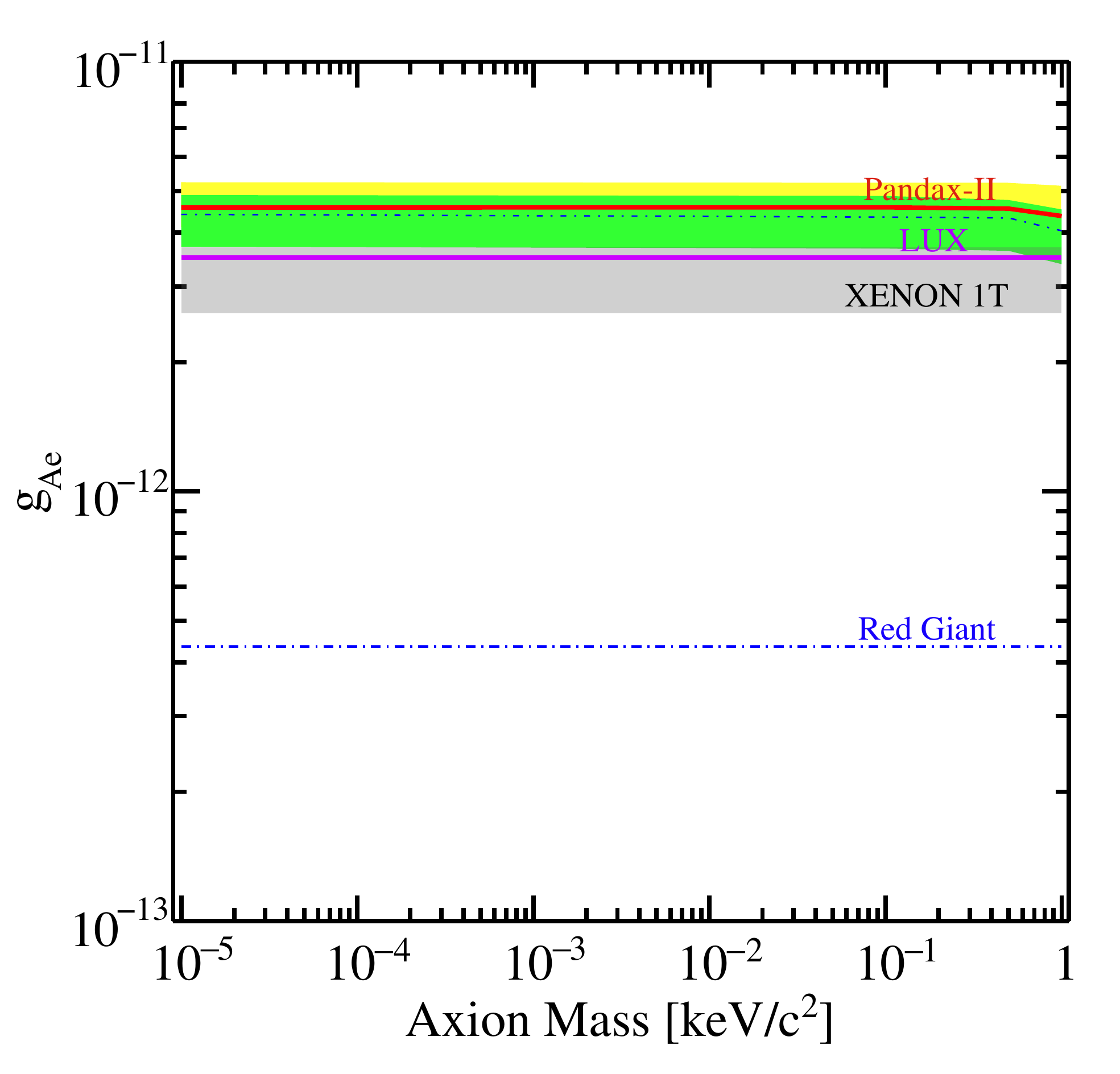}
  \caption{The upper limits on the solar axion coupling constant $g_{Ae}$ (90\% C.L.), overlaid with that from 
  LUX~\cite{LUXaxion}. The best fit region (90\% C.L.) from XENON1T~\cite{Xenon1Texcess} for $g_{A\gamma}<10^{-10}~{\rm GeV^{-1}}$ is shown as a shaded grey region. The green and yellow bands represent the $\pm 1\sigma$ and $ 2\sigma$ sensitivity bands and the dashed line represents the median sensitivity. The upper bounds from red-giant branch observation are also included~\cite{RedGiant}.}
  \label{fig:limit}
\end{figure}

\begin{figure}[H]
  \centering
  \includegraphics[width=0.92\columnwidth]{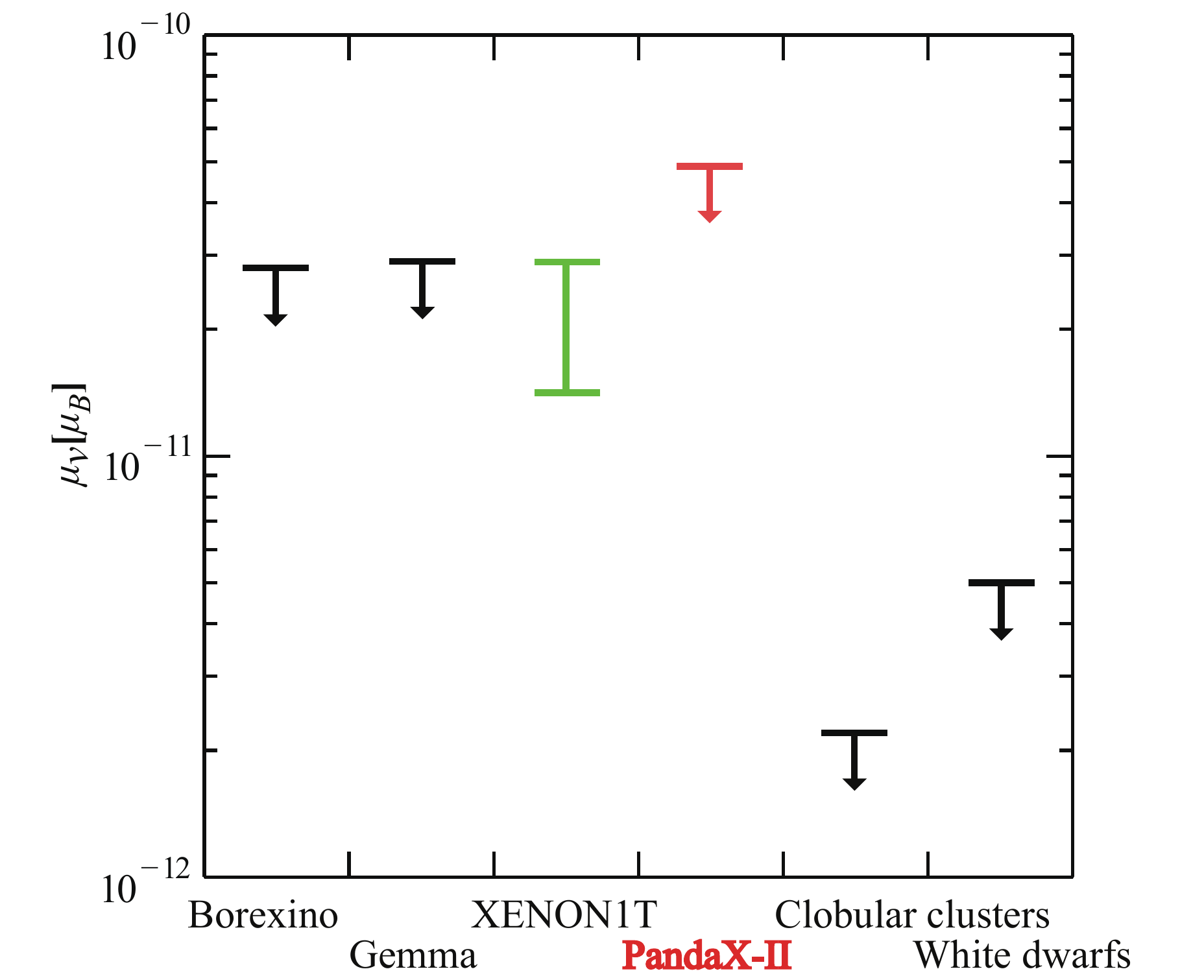}
  \caption{The upper limits on the neutrino magnetic moment (90\% C.L.) is shown in red. The allowed range from XENON1T~\cite{Xenon1Texcess} is marked in green and constraints from other observations~\cite{Borexinodmumag,Gemmamumag,WhiteDwarfmumag, GlobularClustermumag} are shown in black.} 
  \label{fig:mumaglimit}
\end{figure}

To set the exclusion limit, we use the so-called $\rm CL_{s+b}$ method~\cite{CLs} based on profile likelihood ratio~\cite{Cowan:2010js} to make differential comparison of our data with background-only and background-plus-signal hypotheses. The best fit to our data is compared to fits to pseudo-data sets produced at individual signal strength, including statistical fluctuations and spectral shape uncertainties discussed earlier. Constraints on the coupling constant $g_{Ae}$ at 90\% confidence level (C.L.) are shown in Fig.~\ref{fig:limit}. For the axion mass smaller than $0.1~{\rm keV}/c^2$, the upper limit on $g_{Ae}$ is at $4.6\times10^{-12}$, corresponding to 90.9 signal events.
The neutrino magnetic moment hypothesis is tested in the same way, which yields an upper limit of $\mu_\nu$ at $4.9 \times 10^{-11}\mu_B$, corresponding to 191.6 signal events, as shown in Fig.~\ref{fig:mumaglimit}.  
They represent one of the tightest experimental constraints on the solar axion-electron coupling and neutrino magnetic moment. 

To summarize, we perform an analysis using the low energy ER data from the full data set of PandaX-II with the total exposure of 100.7 ton-day. 
Tritium was introduced into PandaX-II during a calibration campaign in 2016 and after the end of physics data taking.
The residual of tritium in the physics data is fitted to be $\rm 0.040\pm 0.010~\mu Bq/kg$, with its shape tightly constrained by calibration. The spectra of $^{85}$Kr and $^{220}$Rn are also extracted from the data directly. With these data-driven background spectra, a search for the solar axion and neutrino magnetic moment signals is carried out. The expected excess assuming the best fit signal strength from XENON1T is compatible with our data within uncertainties, but our data are also consistent with background-only hypothesis.
Upper limits at $90\%$ C.L. on the solar axion and neutrino magnetic moment hypotheses are reported, with $g_{Ae}<4.6\times10^{-12}$ and $\mu_\nu<4.9 \times 10^{-11}\mu_B$. 
The next generation of the PandaX liquid xenon experiment, PandaX-4T~\cite{Zhang:2018xdp}, is expected to lower the electron recoil background rate (per unit target) by more than one order of magnitude, and increase the fiducial volume by about ten times. Together with the upcoming XENONnT~\cite{Aprile:2020vtw} and LZ~\cite{Akerib:2019fml}, a more definitive answer to the  XENON1T excess can be expected in the near future.

\begin{acknowledgments}
This project is supported in part by the Double First Class Plan of the Shanghai Jiao Tong University, grants from National Science Foundation of China (Nos. 11435008, 11455001, 11525522, 11775141 and 11755001), a grant from the Ministry of Science and Technology of China (No. 2016YFA0400301)
and a grant from China Postdoctoral Science Foundation (2018M640036). We thank the Office of Science and Technology, Shanghai Municipal Government (No. 11DZ2260700, No. 16DZ2260200, No. 18JC1410200) and the Key Laboratory for Particle Physics, Astrophysics and Cosmology, Ministry of Education, for important support. We also thank the sponsorship from the Chinese Academy of Sciences Center for Excellence in Particle Physics (CCEPP), Hongwen Foundation in Hong Kong, and Tencent Foundation in China. Finally, we thank the CJPL administration and the Yalong River Hydropower Development Company Ltd. for indispensable logistical support and other help.
\end{acknowledgments}

\bibliographystyle{apsrev4-1}
\bibliography{refs.bib}

\end{document}